\documentclass[12pt]{article}
\usepackage{amsmath,rotating,graphicx,latexsym,verbatim,amssymb,float}
\usepackage[letterpaper, total={6.5in, 9in}]{geometry}
\usepackage{epsfig}
\usepackage{subfigure}
\usepackage{xcolor}
\usepackage{longtable,booktabs}
\usepackage{listings}
\usepackage[square,sort,comma,numbers]{natbib}

\usepackage{lscape} 
\usepackage{graphicx}
\usepackage{caption}

\usepackage{hyperref}
\hypersetup{
    colorlinks,
    linkcolor={blue!50!black},
    citecolor={blue},
    urlcolor={blue!80!black}
   linkbordercolor=blue
}

\renewcommand{\baselinestretch}{1.39}
\makeatletter
\def\singlespace{\def\baselinestretch{0.95}\@normalsize}

\newcommand{\bu}{\mbox{\bf u}}

\newcommand{\bx}{\mbox{\bf x}}

\newtheorem{theorem}{Theorem}
\newtheorem{lemma}{Lemma}

\begin{document}
\setcounter{page}{0}

\begin{center}
{\Large \bf 
 Yakovlev Promotion Time Cure Model with Local Polynomial Estimation}
  \vspace{0.25cm}\\
  {\large {\sc By}  
 {\large \sc LI-HSIANG LIN}}\\
The H. Milton Stewart School of Industrial and Systems Engineering, Georgia Institute of Technology, Atlanta, GA 30332 \\
and   Institute of Statistics, National Tsing Hua University, Hsinchu 30013, Taiwan\\
 {\tt llin79@gatech.edu} \\ 
{\large {\sc AND} \\
  {\large \sc LI-SHAN HUANG}} \\
  Institute of Statistics, National Tsing Hua University, Hsinchu 30013, Taiwan\\
    {\tt lhuang@stat.nthu.edu.tw}\\ 
 \today
\end{center}
\begin{abstract}
\small
 \hspace{-0.62 cm}  
In modeling survival data with a cure fraction, flexible modeling of covariate effects on the probability of cure has important medical implications, which aids investigators in identifying better treatments to cure. This paper studies a semiparametric form  of the Yakovlev promotion time cure model that allows for nonlinear effects of a continuous  covariate. We adopt the local polynomial approach and use the local likelihood criterion to derive nonlinear estimates of covariate effects on cure rates,  assuming that the baseline distribution function  follows a parametric form. This way we adopt a flexible method to estimate the cure rate locally, the important part in cure models, and a convenient way to estimate the baseline function  globally. An algorithm is proposed to implement estimation at both the local and global scales. Asymptotic properties of local polynomial estimates, the nonparametric part,  are investigated in the presence of both censored and cured data, and the parametric part is shown to be root-n consistent. The proposed methods are  illustrated by simulated and real data. 

\end{abstract}


\newpage
\section{Introduction}
Statistical models for survival analysis typically assume that 
every subject in the study is susceptible to relapse if follow-up time 
is sufficiently long. This is often an unstated assumption of 
the widely used Cox's proportional hazards (PH) models.
However, in many clinical studies, we observe that a substantial  portion of 
patients who respond favorably to treatments appear to be free of any symptoms
of the disease and may be considered  ``cured".  
In these cases, investigators observe Kaplan-Meier survival
curves that tend to level off at a value strictly greater than
zero as time increases.  To account for the fact of cure or long-term survivors in practical applications, 
 two classes of cure rate models have been proposed, 
 the two-component mixture (TCM) cure model (Berkson and Gage (1952); Farewell (1982);
Kuk and Chen (1992); Lu and Ying (2004); Mao and Wang (2010), among others) and the Yakovlev promotion time (YPT) cure model (Yakovlev and Tsodikov (1996);  Tsodikov (1998); Chen et al. (1999); Tsodikov et al. (2003); Zeng et al. (2006); Ma and Yin (2008); Bertrand et al. (2017); Chen and Du (2018)),
\begin{align}
    S_{T}(t|X)=\exp(-\theta(X)F(t))\label{PT cure model},
\end{align} 
where $S_T(\cdot)$ is the population survival function for the event time $T$,  $X$ is the covariate and $F(\cdot)$  is an unknown baseline cumulative distribution function. The identifiability of the two types of cure models has been discussed in Li et al. (2001) and Hanin and Huang (2014).  In the literature, the cured or uncured status in the censored set is typically assumed to be not distinguishable for the TCM model, while, in the YPT model (\ref{PT cure model}), if some cured observations are observed, then their  survival time is set as infinity (the event of interest never occurs).  Comparing with the TCM cure model,   the YPT cure model has a natural  biological  motivation (Yakovlev and Tsodikov (1996)) and possesses a population PH structure as Cox's model, which is a desirable property for  survival  models. Chen et al. (1999) discussed some advantages of the YPT cure model. In this paper, we study a local polynomial (Fan and Gijbels (1996)) approach 
of estimating $\theta(\cdot)$ for flexible modeling  of the cure rates under the YPT model. 
To the best of our knowledge, our method is the first to adopt the local polynomial approach for YPT models, while
Chen and Du (2018) take a smoothing spline approach. 

Since  the cure rate under (\ref{PT cure model}),  
\begin{equation}
 \lim_{t \rightarrow \infty} S_{T}(t|X) = \exp(-\theta(X)),
\label{curep}
\end{equation}
 is irrelevant to $F(\cdot)$, we suggest putting more efforts in estimating $\theta(\cdot)$ rather than $F(\cdot)$.  In the literature, some  papers, e.g., Zeng et al. (2006) and Ma and Yin (2008), discussed a semiparametric form for
 (\ref{PT cure model}) by assuming a parametric form for $\theta(\cdot)$   and a nonparametric form
(right-continuous function with jumps)  for
$F(\cdot)$, whereby there may be more efforts in estimating $F(\cdot)$, considered as
  a nuisance parameter in some papers, e.g. Chen and Du (2018).  
We are interested in estimating $\theta(\cdot)$ in (\ref{PT cure model})  nonparametrically for   a  continuous covariate $X$ by the local polynomial approach  (Fan and Gijbels (1996)),
\begin{align}
   \theta(X) = \exp(m(X)),\label{p non}
\end{align}
where $m(\cdot)$ is an unknown smooth function, while assuming a parametric form in estimating $F(\cdot)$. 
Another reason for taking a parametric  $F(\cdot)$ is that it ensures that $m(\cdot)$ is  identifiable (Proposition 7 in Hanin and Huang (2014)).
There are many advantages of modeling $\theta(\cdot)$ nonparametrically, such as offering a more flexible interpretation on the covariate effects for the cure rates.  
Except for the recent publications Lin (2015) and Chen and Du (2018), we are not  aware of papers which adopt a nonparametric approach (\ref{p non}) in YPT cure models. In contrast, there is a rich literature on studying Cox's PH model with nonlinear covariate effects (Tibshirani and Hastie (1987);  Fan et al. (1997); Huang (1999);  Cai et al. (2007); among others) based on the partial likelihood function (Cox (1975)).  Wang et al. (2012) study the TCM cure model with a nonparametric form in the cure probability based on  splines.

     The rest of the article is organized as follows. Section 2 discusses local likelihood under (\ref{PT cure model}) and (\ref{p non}), proposes an algorithm for estimation, and presents the asymptotic properties of estimates  in Theorems 1-3. For estimating $m(\cdot)$ in  
 (\ref{p non}),  Theorem 1 gives the consistency results, and  Theorem 2 shows that the pointwise bias and variance  have the same orders as those  in  Fan et al. (1997) for Cox's model.  The two theorems extend the asymptotic results of local polynomial estimators to cases with  both censored and cured data in survival analysis, while the results from Fan et al. (1997) are for data with censoring.  Theorem 3 investigates the asymptotic normality of the  parameter estimate in $F(\cdot)$. In Section 3,  we use simulated data to examine the performance of the proposed  estimates in practice.  Then the proposed methods are applied to one real data set in Section 4. Some concluding remarks are given in Section 5 and the Appendix provides more technical details for the proofs of the theorems.

\section{Methodology}
\subsection{Estimation}

Consider $n$ independent observations with right censored scheme, $\{(Y_{i} \equiv \min(T_{i},C_{i}),\Delta_{i} \equiv I(T_{i} < C_{i}),X_{i}): i = 1,\cdots,n\}$, where $T_{i}$, $C_{i}$, and $X_{i}$ are the failure time, censoring time, and covariate for the $i$-th observation, respectively, and $I(\cdot)$ is the indicator function. Furthermore,  $T_i$ is conditionally independent of $C_i$ given $X_i$. We assume that the follow-up time of cured subjects is infinite, theoretically, and that a proportion of subjects is cured without experiencing failure or right censoring $(T_i=C_i=\infty)$, i.e. $P(Y=\infty |X) >0$.
In practice,	to claim a subject is cured or not, a cure threshold $\zeta$ may be defined. 
For the observations with $Y_{i} \leq \zeta$, they are  either a failure $(\Delta_{i}=1)$ or right-censored $(\Delta_{i}=0)$. Those observations with long survival time $Y_{i} >\zeta$ are classified as cured. Since cured subjects never experience the failure, their $T_i$'s and $C_i$'s are set as $\infty.$ 
This approach of using the cure threshold  to identify cured observations has been adopted in some papers, e.g.,  Zeng et al. (2006) and Ma and Ying (2008), since an infinite follow-up time is not practical.  Note that for those subjects with $Y_i <\zeta$ and $\Delta_i=0$, their final cure/failure status is unknown. 

 For estimating $m(\cdot)$ nonparametrically in (\ref{p non}), $X$ is assumed to be a univariate continuous covariate. Under model (\ref{PT cure model}), the population hazard function is 
$h_p(t|X)= \theta(X) f(t)$ (Chen et al., 1999), where $f(\cdot)$ is the density function of $F(\cdot)$, and hence the cumulative hazard function is $\theta(X) F(t)$.  When $\theta(X)$ takes a parametric form $\exp(b_0+ b_1X)$,  $\exp(b_0) F(t)$ is the cumulative baseline hazard function (Zeng et al., 2006). Since we choose a nonparametric form   (\ref{p non}),  the cumulative baseline hazard function in this case is $\exp(m(0)) F(t)$. However, the range of $X$ does not necessarily include  0, and hence $\exp(m(0))$, which is the upper bound of the cumulative baseline hazard function, may not be estimated. Therefore   $F(t)$ may be interpreted as the cumulative baseline hazard function subject to a possibly unknown  constant multiple.

Let $\gamma$ denote the parameter of $F(\cdot)$ in (\ref{PT cure model}).
 Then the likelihood function given  $(X_1, \dots, X_n)^T$ is
\[
   L= \prod^{n}_{i=1}\lbrace [f_{T}(Y_{i};\theta(X_{i}),\gamma)]^{\Delta_{i}}[S_{T}(Y_{i};\theta(X_{i}),\gamma)]^{1-\Delta_{i}}\rbrace^{I(Y_{i} < \infty)}\lbrace S_{T}(\infty; \theta(X_{i}),\gamma)\rbrace^{I(Y_{i} = \infty)}.
\]
See, e.g., Ma and Yin (2008), for derivations of $L$ under (\ref{PT cure model}).
The corresponding log likelihood function   is
  \begin{align}
\ell =  \sum^{n}_{i=1} \left\{ \Delta_{i}[\log(\theta(X_{i})) + \log f(Y_{i}; \gamma)] - \theta(X_{i}) F(Y_{i}; \gamma) \right\}, \label{log likelihood fun}
 \end{align}
and  the derivation of (\ref{log likelihood fun}) is given in the Appendix. Note that $F(\infty; \gamma)=1$ for those cured observations with $Y_i=\infty$.    In this paper, we are interested in estimating $\theta(\cdot)$ in a nonparametric form (\ref{p non}), while assuming $F$ in a parametric form. This way, we adopt a flexible method to estimate cure rates, the important part in cure models, and a convenient way to estimate the baseline $F(\cdot)$, which is not involved in  (\ref{curep}). 

On  the cure threshold $\zeta$, some papers use the largest failure time as the cure threshold (Laska and Meisner (1992), Zeng et al. (2006), and Ma and Yin  (2008)). Since the threshold for cure implies medical decisions, it may be   determined by
physicians in practice. From a statistical point of view, we may re-fit the model using
different cure thresholds. In section 4, we illustrate the effects of using different cure thresholds on the estimation of cure rates by real data analysis. 

The following lemma shows that $\theta(x)$ can  be estimated if $F(\cdot)$ is known.
\begin{lemma}
 Under model (\ref{PT cure model}) with (\ref{p non}) and conditions (A8) in the Appendix, denote the true value  of $\gamma$ as $\gamma_{0}$. Then
\begin{equation}
\theta(x) = \frac{E[\Delta|X=x]}{E[F(Y;\gamma_{0})|X=x]}.
\label{lemma1}
\end{equation}
 \end{lemma}
The proof of Lemma 1 is given in the Appendix. 
Results similar to Lemma 1 for  
 Cox's model can be found in Fan et al. (1997), p:1662, where  the denominator of (\ref{lemma1}) is replaced by the expected cumulative baseline hazard function under Cox's model.

With a nonparametric form  (\ref{p non}) for  $\theta(X)$,  as in  Fan and Gijbels (1996), based on a Taylor expansion of $m(X_{i})$ at a grid point $x$, 
 $m(X_{i})$ is approximated  by  $m(x) + m^{'}(x)(X_{i}-x)+(m^{''}(x)/2!)(X_{i}-x)^{2} + \hdots + (m^{(p)}(x)/p!)(X_{i}-x)^{p}$  for $X_{i}$ in a neighborhood of $x$. 
  Then incorporating local weights  around $x$ for $\ell$,  a local likelihood at $x$ is
   \begin{align}
\ell_{x}(\beta_{x}; \gamma)= n^{-1} \sum^{n}_{i=1}\{\Delta_{i}[\tilde{ \mathbf{x}}_{i}^{T}\beta_{x}+\log f(Y_{i};\gamma)]-\exp(\tilde{ \mathbf{x}}_{i}^{T}\beta_{x})F(Y_{i};\gamma)\}K_{h}(X_{i}-x),
\label{local log-like}
 \end{align}
where $\beta_{x} = (\beta_{0},\hdots,\beta_{p})^{T}=( m(x), m^{'}(x),...,m^{(p)}(x)/p! )^{T}$,  $ \tilde{ \mathbf{x}}_{i}=(1,X_{i}-x,\hdots,(X_{i}-x)^{p})^{T},$
and $K_{h}(\cdot)=h^{-1} K(\cdot/h)$ with $K$ the kernel function and $h$ the bandwidth. 
The local parameters  $\hat{\beta}_{x}$'s are estimated by  maximizing the local likelihood (\ref{local log-like}) for a set of grid points $x$'s and  estimates of $\beta_0(X_i)$'s are typically used for estimating $m(X_i)$'s. The $n^{-1}$ factor  in (\ref{local log-like}) is included for technical convenience
when  deriving the asymptotic results in Theorems 1, 2, and 3 of this paper, analogous to Fan et al. (1997)  for Cox's model, and it does not affect the maximization of the local likelihood.

The concavity of the local likelihood function (\ref{local log-like}) is shown in the following Lemma.
\begin{lemma}
Under conditions (A)  in the Appendix, given a fixed $\gamma$, the local likelihood (\ref{local log-like})  is strictly concave down
and hence the maximizer to  (\ref{local log-like})  is unique. 
\end{lemma}
Lemma 2 is shown by taking the Hessian matrix of  $\ell_{x}(\beta_{x}; \gamma)$ with respect to $\beta_{x}$, 
$$  \frac{1}{n} \sum^{n}_{i=1}\{-\exp(\tilde{ \mathbf{x}}_{i}^T \beta_{x})F(Y_{i};\gamma)\}
\tilde{ \mathbf{x}}_{i} \tilde{ \mathbf{x}}_{i}^T
K_{h}(X_{i}-x),$$
which is negative definite
since   $F(\cdot)$,  $\exp(\cdot),\ \mbox{and}\ K(\cdot) > 0$.

We note that the cured subjects
with $F(\infty)=1$ do not contribute information in  estimating $\gamma$. 
Thus estimating $\gamma$ may be based on maximizing the conditional likelihood for  failure and censored subjects only, 
conditioned on the fact that  their  $Y_{i}$'s $< \infty$: 
\begin{eqnarray}
\ell^* & = & n^{-1}  \sum^{n}_{i=1}  I(Y_{i} < \infty)   \left\{  \Delta_{i} [\log(\theta(X_{i})) + \log f(Y_{i}; \gamma) - \theta(X_{i}) F(Y_{i}; \gamma)] \nonumber  \right. \\
 & & \left. + (1-\Delta_{i}) [ \log ( \exp(-\theta(X_{i}) F(Y_{i}; \gamma)) - \exp(-\theta(X_i)))] - 
(1- \exp(-\theta(X_i) )) 
 \right\}.
\label{noncured}
 \end{eqnarray}
Based on our experience,  maximizing the conditional likelihood (\ref{noncured}) for estimating $\gamma$  has better numerical performance than that without conditioning.
Then estimates of  $\beta_x$  and $\gamma$ may be obtained by iteratively  maximizing 
 local likelihood  (\ref{local log-like}) and conditional likelihood (\ref{noncured}).  The iterating algorithm is stated as follows.
\begin{itemize}
\item [1.] Given the observed cure rate $\bar{p}=\#\{Y_{i} > \zeta\}/n > 0,$ 
set an initial value of  $\tilde{\beta}^{(0)}_{0}(\cdot)= \log( -\log \bar{p})$. 
Maximize  (\ref{noncured})  with respect to $\gamma$ to obtain 
an initial value $\tilde{\gamma}^{(0)}$.
\item [2.]  With  $\tilde{\gamma}^{(r)}$, use a small bandwidth (to be explained in Theorem 3)  in maximizing
 (\ref{local log-like}) to obtain estimates $\tilde{\beta}_{x}^{(r)}$ for a set of grid points that include data points $X_{1}, \dots, X_n$, and then  $\tilde{\beta}_0^{(r)}(X_i)$ estimates $m(X_i)$, $i=1, \dots, n$. Based on  (\ref{p non}), calculate $\tilde{\theta}^{(r)}(X_i)$,  $i=1, \dots, n$.
\item [3.] With $\tilde{\theta}^{(r)}(X_i)$'s, maximize  (\ref{noncured})  with respect to $\gamma$ to update 
 $\tilde{\gamma}^{(r+1)}$.
\item [4.] Iterate steps 2 and 3 until some convergence criterion is satisfied; for example, both $|\tilde{\gamma}^{(r+1)}-\tilde{\gamma}^{(r)}|$ and   $\max_{i}|\tilde{\theta}^{(r+1)}(X_{i})-\tilde{\theta}^{(r)}(X_{i})|$ are less than $10^{-4}$. 
Denote the final estimate for $\gamma$ as   $\hat{\gamma}$.
\item [5.] Take the bandwidth suitable for $m(\cdot)$ and re-estimate $m(\cdot)$ in step 2 with $\hat{\gamma}$.  Estimates of $\theta(X_i)$'s are then obtained by (\ref{p non}) and the estimated cure rates by (\ref{curep}).
\end{itemize}

\subsection{Asymptotic Properties}
The following two theorems describe the asymptotic properties of $\hat{\beta}_x$ and their proofs are given in the Appendix.

\begin{theorem}
Under conditions (A) in the Appendix, given true $\gamma = \gamma_{0}$,
$\hat{\beta}_{x}$ is a consistent estimator of true $\beta^{0}$ in the sense that
$$H(\hat{\beta}_{x} -  \beta^{0})  \stackrel{p}{\rightarrow} 0,$$
where  $H$ is a diagonal matrix with entries $(1, h, \dots, h^p)$.
\end{theorem}

\begin{theorem}
Under conditions (A) in the Appendix, (a) given true $\gamma = \gamma_{0}$ and the local polynomial order $p$ is odd, 
$$\sqrt{nh}\left( H(\hat{\beta_{x}}-\beta_{x}) - b_n(x) \right) \rightarrow N_{p+1}(0, f_X(x)^{-1}S^{-1}_{1}(x)S_{2}(x; \gamma_{0})S^{-1}_{1}(x)),$$
where  $f_X(\cdot)$ is the marginal density function of $X$, 
\begin{equation}
b_n(x)= \frac{m^{(p+1)}(x)}{(p+1)!}h^{p+1} \left(\int  \begin{bf}uu^{T}\end{bf}K(u)du \right)^{-1} \int  \begin{bf}u\end{bf}u^{p+1}K(u)du + o_{p}(h^{p+1})
\label{bias}
\end{equation}
with \begin{bf}u\end{bf} $= (1, u, \hdots, u^{p})^T$, 
 $ S_{1}(x) =   E(\Delta |X=x) \int  \begin{bf}uu^{T}\end{bf}K(u)du$, and $ S_{2}(x;\gamma_{0}) = E(([\Delta - \theta(x)]F(Y;\gamma_{0}))^{2}|X=x)\int 
\begin{bf}u\end{bf}\begin{bf}u\end{bf}^{T} K^{2}(u)du$;\\
(b) When $\gamma$ is unknown and estimated by a $\sqrt{n}$-consistent $\hat{\gamma}$, results in (a) continue to hold.
\end{theorem}
Theorem 2 shows that  the leading terms for the bias and variance of $\hat{\beta}_{0}$ are of order  $h^{p+1}$ and $(nh)^{-1}$ respectively, which are similar to the estimators of local polynomial regression (Fan and Gijbels, 1996) and to maximum partial likelihood estimators for Cox's model (Fan et al., 1997).  In addition, the variance of $\hat{\beta}_x$ depends on the censoring scheme  and the baseline  $F(\cdot)$, while the bias  comes from the approximation error.   In practice, the variance of $\hat{\beta}_x$ may be empirically estimated
by the inverse of the observed local information matrix.
In Theorem 2(b),  when   $\hat{\gamma}$ is root-$n$ consistent, the nonparametric component is estimated  as if  $\gamma$ was known as in Theorem 2(a).
This result is similar to   Cai et al. (2007) under the Cox's model and will be illustrated by simulated data in Section 3.

Based on Theorem $2$, the theoretical optimal bandwidth in terms of minimizing the weighted mean integrated squared error for estimating $m(\cdot)$ is 
$$\frac{n^{\frac{-1}{2p+3}}}{2p+2}\left\lbrace\int v(x;\gamma_{0})w(x)dx\right\rbrace^{\frac{1}{2p+3}}\left\lbrace\int (b^{(1)}_{n}(x))^{2}w(x)dx\right\rbrace^{\frac{-1}{2p+3}},$$
where $v(x;\gamma_{0})$ is the (1,1)-th element  of $f^{-1}_{X}(x)S^{-1}_{1}(x)S_{2}(x;\gamma_{0})S^{-1}_{1}(x)$, $b^{(1)}_{n}(x)$ is the first element of $ m^{(p+1)}(x)/(p+1)! \left(\int  \begin{bf}uu^{T}\end{bf}K(u)du \right)^{-1} \int  \begin{bf}u\end{bf}u^{p+1}K(u)du$ from $b_{n}(x)$ in Theorem 2, and $w(\cdot)$ is a weight function. The  order of bandwidth required for $\sqrt{n}$-consistency of estimating $\gamma$ will be given in Theorem 3,  and developing a data-driven procedure to selecting  the bandwidth will be an interesting topic for future work.     

Since the cure rate $S_T(\infty|X)= \exp(-\exp(m(X)))$ under  (\ref{PT cure model}),   $m(\cdot)$ is estimated with inference  on the cure rates. This implication  is different  from Fan et al. (1997) and Cai et al. (2007)  for  Cox's model, as the interest there is on 
estimating the first derivative for the hazard function. The rates of bandwidth are also different; we focus on an odd $p$ for estimating $m(\cdot)$, while the work of Fan et al. (1997) and Cai et al. (2007) adopt an even $p$ for estimating the first derivative.

Then we  derive the asymptotic bias and variance of $\hat{\gamma}$  and the proof is given in the Appendix.
\begin{theorem}
 Under conditions (A)  in the Appendix, \\
(a) given true $\theta(\cdot)$, 
 $\sqrt{n} (\hat{\gamma} - \gamma_0)$ converges to a Gaussian distribution with mean 0 and variance
$T_{1}(\gamma_{0})^{-1}T_{2}(\gamma_0)T_{1}(\gamma_{0})^{-1},$
where $T_{1}(\gamma_{0})$ and $T_{2}(\gamma_{0})$ are given in (\ref{t1}) and (\ref{t2})  respectively in the Appendix;\\
(b) when $\theta(\cdot)$ is estimated with rates in Theorem 2 and the bandwidth satisfying $nh^{2p+2} \rightarrow 0$ for an odd $p$,   the results in (a)  continue to hold.
\end{theorem}
Theorem 3 shows  that, given true $\theta(\cdot)$, $\hat{\gamma}$ is unbiased and its variance has a parametric rate $1/n$. When $\theta(\cdot)$ is estimated with rates in Theorem 2, a small bandwidth is needed to ensure the root--n rate of $\hat{\gamma}$. For example, for local linear regression $p=1$, the order of $h$,  $n^{-a}$, needs to satisfy  $1/4 < a < 1$, which is smaller than the typical rate $n^{-1/5}$. This result is similar to Huang (1999) and Cai et al. (2007) that  the $\sqrt{n}$-rate of convergence and asymptotic normality of $\hat{\gamma}$ hold for a range of
 the smoothing parameter. A naive and practical approach for estimating the variance of $\hat{\gamma}$ is by taking the inverse of the empirical Fisher information based on the conditional likelihood (\ref{noncured}). This naive approach will be examined numerically in Section 3.
Though the cured subjects do not contribute to  estimation of $\gamma$,  the convergence  rate in Theorem 3 retains the root-n rate. This is due to the fact that the cure rate under model (\ref{PT cure model}) does not depend on $n$, and it only affects the constant terms $T_{1}(\gamma_{0})$ and $T_{2}(\gamma_{0})$ (see (\ref{t1}) and (\ref{t2}) in the Appendix with an indicator function $I(Y < \infty)$).

\section{Simulation Study}
In this section we evaluate the performance of the proposed estimation approach by two simulated examples. The sample size is $n=$200 and the number of simulations is 1000 in each example.  The data generation scheme for an improper $S_{T}(\cdot|x)$ (\ref{PT cure model}) is described as follows.
The covariates $X_i$, $i=1, \dots, 200$ are generated independently from  a given distribution  $F_{X}(\cdot)$ of $X$, and for a given function  $m(\cdot)$,  $\theta(X_i)= \exp(m(X_i))$ (\ref{p non}), $i=1, \dots, n$,  are obtained. The true cure rates are calculated  by $p_i=\exp(-\theta(X_i))$ (\ref{curep}). Then  independent uniform values $U_i$'s on $(0,1)$  are generated. If $U_i< p_i$, then the $i$-th observation is cured with $Y_i=T_i=C_i=\infty$; otherwise  $Y_i < \infty$.  For $Y_i < \infty$, with a given baseline $F(\cdot)$, we find $T_i$ such that   $U_i= S_{T}(T_i|X_i)=\exp(-\theta(X_i) F(T_i))$ and then  $T_i$ is compared with  censoring time $C_i$, which is independently generated. If $T_i < C_i$, then $Y_i=T_i$ and $\Delta_i=1$; otherwise $Y_i=C_i$ and $\Delta_i=0$. 
 Note that in the Cox's model setting,  the  cured observations were  classified as censored. Following this convention, we adopt the term  ``the overall censoring rate" for the proportion of subjects with 
   $(Y < \infty$ and $\Delta=0)$ or with $Y= \infty$. Thus the overall censoring rate is greater than or equal to the cure rate.

   We use local linear regression $p=1$ with the Epanechnikov kernel function for smoothing. The continuous covariate $X \sim U(1, 4)$, and 
 for estimating  $m(\cdot)$,  301 equally-spaced grid points  in the  support (1, 4) are used to estimate the curves. The average mean squared error (MSE) is calculated  for interior grid points on [1.3, 3.7] to avoid boundary effects.
  
\noindent {\bf Example 1: $\theta(X) = \exp(1 + \sin(2X))$}\\
This function and its cure rate  (solid lines in Figure  1(a)(b)  respectively)  look like a quadratic function visually. The baseline  $F(\cdot)$ is taken as an exponential distribution with parameter  $\gamma = 7$  and the censoring time $C  \sim U(0,1)$.  The resulting overall censoring and cure rates are $19\%$ and $13.5\%$, respectively, which means $5.5\%$ are censored but not cured. 
We first try using the same  bandwidth $h=$0.2, 0.4, and 0.6  in steps 2 and 5 in the proposed algorithm. The mean  estimates  (sd) for $\gamma$ are given in Table 1 as well as the average of the estimated standard error ($\widehat{se}$) by the inverse of observed Fisher information. The  coverage rate of 95$\%$ confidence intervals for $\hat{\gamma}$ is also computed. 
It is seen that $\hat{\gamma}$ is close to true value,  $\widehat{se}$ is close to the sd of $\hat{\gamma}$, and the average coverage rate is  reasonably close to the nominal level among the 1000 simulations. In this example, the estimation of $\gamma$ seems not sensitive to the values of the bandwidth, possibly due to a low censoring rate. When $\gamma$ is estimated and $h=0.6$,  Figure 1(a)  shows the estimated functions  $\hat{m}(\cdot)$ with performance at 10-, 50-, and 90-th percentiles of the average MSE over interior grid points among the 1000 simulations, with the corresponding cure rates in Figure 1(b).  The performance of $h=0.2$ and $0.4$ (not shown)  is similar to that of $h=0.6$. It is evident that the proposed estimation performs reasonably well even when the same bandwidth is used in estimating $\gamma$ and $m(\cdot)$ in this example. We also evaluate estimation of $m(\cdot)$ when $\gamma$ is known vs. estimated (unknown). Table 1 includes the resulting average MSEs and we observe that estimation of $\gamma$ slightly affects estimation of $m(\cdot)$. In addition,    $\hat{m}$ has a smaller MSE when $h=0.6$. To assess the sampling variability of $\hat{m}(\cdot)$ at each grid point, for the estimated function with 50-th percentile of average MSE, the inverse of observed local Fisher information matrix is calculated  with $h=0.6$ and the resulting 95\% pointwise confidence intervals  are illustrated in Figure 1(c) ($\gamma$ known) and 1(d) ($\gamma$ unknown).  The results show that the nonparametric function is estimated with reasonable accuracy and is not heavily dependent on the parametric  part. The 95$\%$ pointwise confidence intervals for $h=0.2$ and 0.4 (not shown) are slightly wider than that of $h=0.6$, as the pointwise variance has an order of $(nh)^{-1}$ (Theorem 2).

\noindent {\bf Example 2: $\theta(X) = \exp( \sin(2X))$}\\
This function is similar to that of Example 1 except  that  the intercept is 0, in order to increase the cure rate. The baseline  $F(\cdot)$   and the censoring time $C $ are the same as in Example 1.  The resulting censoring and cure rates are 44.6$\%$ and 38.6$\%$, respectively; as a result, $6\%$ observations are censored but not cured. We want to examine whether the performance seen in Example 1 is affected after increasing the cure rate.
Again we  try using the same  bandwidth $h=$0.2, 0.4, and 0.6  in steps 2 and 5 in the proposed algorithm.
Table 2 describes the performance of  $\hat{\gamma}$, and it is seen that 
estimation of $\gamma$ continues to perform  well, not sensitive to the values of the bandwidth. We note that
the average of  $\widehat{se}$  is larger than that of Example 1, possibly due to a higher cure rate (recall that the cured subjects do not contribute to estimation of $\gamma$).
Table 2 also gives the average MSE of $\hat{m}(\cdot)$ when $\gamma$ is known vs. estimated, 
 and $\hat{m}$ has  larger MSEs as compared to those in Example 1, especially when $h=0.2$. Thus we may infer that a higher cure rate affects  estimation of both $m(\cdot)$ and  $\gamma$.  Figure 2 shows that the estimated functions and their cure rates with $h=0.6$ are visually similar to those in Figure  1, and the cure rates in Figure 2(b) are higher than  those in Figure 1(b). The 95\% pointwise confidence intervals in Figure 2(c)(d)  for the estimated function with 50-th percentile of average MSE, are slightly wider than those of Figure 1(c)(d), possibly due to a higher cure rate.


%

  \noindent {\bf Example 3: }\\
This example is the same as  Example 1 except  that   the censoring time $C \sim  U(0, 0.4)$, to increase the censoring rate.  In this example, the censoring and cure rates are 26.8$\%$ and 13.5$\%$, respectively, which indicates $13.3\%$ observations are censored but not cured. We want to examine whether the performance seen in Example 1 is affected after increasing the censoring rate.
We first  try  the   bandwidth $h=$0.2, 0.4, and 0.6  in step 2  of the proposed algorithm and find that
the average $\hat{\gamma}$'s are  7.56  and 7.60  when $h= 0.4$ and 0.6 respectively, which show a sizable difference from the true value 7. Hence we choose a smaller bandwidth $h=0.2$  for estimating $\gamma$ in step 2 and the average $\hat{\gamma}$ (sd) is 7.293 (1.049), which is not as good as in Examples 1 and 2. The corresponding $\widehat{se}$ and the average  coverage probability are 1.398 and 96$\%$ respectively.
Then $h=0.4$ and 0.6  are used in step 5 for estimating $m(\cdot)$ and the average MSE (sd)   are 0.062 (0.043) and 0.041 (0.032) respectively when $\gamma$ is estimated. 
It is seen that $\hat{m}$ has  larger MSEs generally  as compared to those in Example 1, and in step 5, using $h=0.6$ has a smaller average MSE than that of $h=0.4$.  The estimated functions and their cure rates  (not shown)  were examined. Overall, the estimated curves exhibit more variability than  those of Examples 1.
From this example, it is seen that increasing  the censoring rate affects the choice of $h$ and   estimation of both $m(\cdot)$  and $\gamma$. 

\section{Analysis of Kidney Transplant Data}    

    This dataset consists of 863  subjects with 140 failures and  originally 723 censored from Klein and Moeschberger (1997). The endpoint is  the time to death or censored for kidney transplant patients who had their transplant performed at the Ohio State University transplant center during  1982-1992. The maximum follow-up time  was  3434 days (9.47 years). Censored information occurred if they moved from Columbus (lost to follow-up) or if they were alive on June 30, 1992. The only continuous covariate in the data is patient age, range [1, 75] years and mean 42.84 (sd 13.52) years, and discrete covariates include gender (males 60.7\%, females 39.3\%) and race
(whites 82.5\%; blacks 17.5\%).
   The largest failure time is  3147 days, and we choose 3147  as the cure threshold,  which leads to 37(4.3\%) cured and 686(79.5\%) censored.  Among the cured patients, their age has a range [6, 61] years and mean 35.30 (sd 12.04) years, the range of on-study time [3161, 3454] days, 18 males and 19 females,  and 31 whites and 6 blacks. Klein and Moeschberger (1997) used a kernel smoothing procedure to estimate the hazard rate and this motivates us to apply our  methods to this dataset for a comparison.

The proposed cure model with bandwidth $h=10$ and 12 is used to implement step 2 in the algorithm, and the resulting $\hat{\gamma}$ is   the same to the 6th decimal places, 8.4$\times 10^{-5}$. 
Since $F(\cdot)$ is the baseline cumulative hazard function subject to a constant multiple,
 the interpretations for the small $\hat{\gamma}$ may be as follows:  the  baseline kidney transplant failure 
 tends to happen more slowly,
with estimated mean 
  about 11905 days (32.6 years), and median 8251 days (22.6 years).
 Then in step 5, $h=18$ and 22 are used,  and  the resulting $\hat{m}(x)$ and  estimated cure rates with 95$\%$ confidence interval   are shown  in Figure \ref{R2}. The trend is monotonic, though not exactly linear; the younger the patient age, the higher the cure probability. When $h = 10$ in step $2$, the estimated $\hat{se}$ of $\hat{\gamma}$ with $h=18$ and $h=22$ in step 5 are 1.21$\times 10^{-5}$ and 1.20$\times 10^{-5}$  respectively. The performance of $\hat{\gamma}$ for  $h = 12$ in step 2 is similar to that of $h=10$.
 
  We also re-fit the proposed methods  using different cure thresholds 3100, 3200, and 3300 days based on $h=10$ in step 2 and  $h=22$ in step 5, and  $\hat{\gamma}$ is $8.9\times 10^{-5}$, $8.0 \times 10^{-5}$, and $7.4\times 10^{-5}$ respectively ($\hat{se}=1.1\times 10^{-5}, 1.0\times 10^{-5},\mbox{and } 9.0 \times 10^{-6},$ respectively).  The resulting estimated cure rate curves  with their $95\%$ confidence intervals  are  plotted in Figure  \ref{R2_diffc}(a)(b).  It shows that the higher the cure threshold, the lower the estimated cure rate. To empirically check for goodness of fit of the  fitted model, we plot the estimated survival curve and the Kaplan-Meier curve in Figure  \ref{R2_diffc}(c), where  the estimated survival curve is the empirical average of the estimated survival functions with follow-up time ranging from 0 to the cure threshold 3147. The  plot indicates that the proposed model fits the dataset reasonably well.

     Klein and Moeschberger (1997) p:171  shows a kernel smoothed hazard rate based on the Nelson-Aalen estimator.  Figure  \ref{R2_diffc}(d)  plots the estimated hazard functions  by the proposed methods  conditioned on  patient age 33 (first quartile), 42.84 (mean), and 54 (3rd quartile) years. We observe that the hazard functions under (\ref{PT cure model}) are bounded, while  the  estimated hazard function  shown in Klein and Moeschberger (1997) is not bounded,  which is a key difference between Cox's and YPT  models.    

\section{Discussion}
Cure rate models have been shown to be useful for analyzing time-to-event data and they provide different
interpretations from  conventional Cox's models. In this paper, we explore estimating nonlinear covariate effects for the YPT cure model based on local polynomials and retaining a parametric baseline $F(\cdot)$. With the presence of both cured and censored data, our results show that the nonparametric part can be estimated with typical rates as in Fan and Gijbels (1996) and the parametric part can be estimated at a root-n rate. The proposed methods are limited to the case with one continuous covariate and a future extension is on accommodating both linear and nonlinear covariate effects, such as the partially linear structure, for the YPT model. Such model structure also allows estimating cure rates with both qualitative and quantitative covariates. We conjecture that the proposed methodology continues to apply, as there is a global parameter for $F(\cdot)$ and the partially linear structure also includes global parameters, but with more techniques involved.

In the literature on the YPT cure model, a group of cured subjects is assumed to be observed based on  physicians' judgments or diagnostic procedures in clinical studies. However, we may only observe some evidence of long-term survivors in practice but do not know whether they are cured. For such a situation,  to distinguish cured from  censored subjects, a common approach is to define a cure threshold which may be the largest failure time. To our  knowledge, we are not aware of works on estimating the cure threshold based on some statistical criterions.  It will be interesting to develop some statistical methods to estimate the cure threshold for cure models. In addition, exploring the effects of different cure thresholds on estimating cure rates may be  an interesting future research topic.

\section{Appendix} 

\begin{bf}
 Derivation of the  log likelihood function (\ref{log likelihood fun}):
\end{bf}
\begin{align}
\ell & = \log \left( \prod^{n}_{i=1}\lbrace [f_{T}(Y_{i};\theta(X_{i}),\gamma)]^{\Delta_{i}}[S_{T}(Y_{i};\theta(X_{i}),\gamma)]^{1-\Delta_{i}}\rbrace^{I(Y_{i} < \infty)}\lbrace S_{T}(\infty; \theta(X_{i}),\gamma)\rbrace^{I(Y_{i} = \infty)} \right)\nonumber\\
 & = \sum^{n}_{i=1} I(Y_{i} < \infty)\left\{\Delta_{i}[\log(\theta(X_{i})) + \log f(Y_{i}; \gamma) - \theta(X_{i}) F(Y_{i}; \gamma)]  - (1 - \Delta_{i})\theta(X_{i}) F(Y_{i}; \gamma)\right\rbrace - \nonumber\\
 & \hspace{1 cm}  I(Y_{i} = \infty)[\theta(X_{i}) F(Y_{i}; \gamma)]\nonumber\\
  & = \sum^{n}_{i=1} I(Y_{i} < \infty)\left\{\Delta_{i}[\log(\theta(X_{i})) + \log f(Y_{i}; \gamma)]  - \theta(X_{i}) F(Y_{i}; \gamma)\right\rbrace -  I(Y_{i} = \infty)[\theta(X_{i}) F(Y_{i}; \gamma)]\nonumber\\
& = \sum^{n}_{i=1} \left\{\Delta_{i}[\log(\theta(X_{i})) + \log f(Y_{i}; \gamma)]  - \theta(X_{i}) F(Y_{i}; \gamma) \right\} 
\end{align}
\noindent \textbf{Conditions (A):}
\begin{enumerate}
\item [(A1)]The kernel function $K(\cdot) \geq 0$ is a bounded density with compact support. 
\item [(A2)]The function $m(\cdot)$ has a continuous $(p+1)$-th derivative around the point $x$.
\item [(A3)]The second derivative of the baseline function $F(t)$ exists.
\item [(A4)]The density of $X$, $f_X(x) >0$,  is continuous.
\item [(A5)]For estimating the $k$-th derivative of $m(\cdot)$, $k=0,  \dots, p$, $h \rightarrow 0$ and $nh^{2k+1} \rightarrow \infty$, as $n \rightarrow \infty$. 
\item [(A6)] The true value of $\gamma$, $\gamma_0$, in  the baseline function $F(\cdot)$ is an interior point of its parameter space.
\item [(A7)]There exists an $\eta > 0$ such that $E\{|\Delta - \theta(x)F(Y;\gamma_{0})|^{2+\eta}\}$ are finite and continuous at the point $X=x$ for $0<\eta<1$.
\item [(A8)]The functions $E\{\Delta|X\}$,  $E\{F(Y; \gamma_{0})|X\}, E\{F^{'}(Y; \gamma_{0})|X\}$, $E\{F^{''}(Y; \gamma_{0})|X\}, E\{\xi(Y; \gamma_{0})|X\}$, $E\{\xi^{'}(Y; \gamma_{0})|X\},$ and $E\{\xi^{''}(Y;\gamma_{0})|X\}$ are continuous at point $X=x$, where   $\xi(Y; \gamma) = \log f(Y; \gamma)$.
\item [(A9)] There exists a function $M(y)$ with $EM(Y) < \infty$ such that 
$$ \left| \frac{\partial^{3}}{\partial \beta_{j}\partial \beta_{k}\partial \beta_{l}} \ell_{x}(\beta) \right| < M(y).$$
\end{enumerate}
\begin{bf}
Proof of Lemma 1\\
\end{bf}
The log likelihood function   (\ref{log likelihood fun})  is the data version of 
the population  log likelihood function  
$$ \ell_1=  \Delta(\log\theta(X)+\log f(Y; \gamma)) - \theta(X)F(Y;\gamma).$$
With $\gamma=\gamma_0$,  at point $x$, 
taking the derivative of $\ell_1$ with respect to $\theta(\cdot)$  and then taking the conditional expectation,
\begin{eqnarray}
 E\left(\frac{\partial \ell_1 }{\partial \theta(x)}|X = x \right) = E\left(\Delta( 1/ \theta(x)) -  F(Y; \gamma_{0})|X = x \right), \nonumber 
\end{eqnarray}
which is 0 at  the true value of $\theta(x)$. Thus Lemma 1 is obtained.\\

\begin{bf}
\noindent Proof of Theorem 1\\
\end{bf}
For simplicity, the subscript of $\beta_x$ is neglected in this proof.  Recall that 
$\beta^{0}$  is the  true value  of $\beta$.
Given $\gamma = \gamma_{0}$,  let $\alpha = H(\beta - \beta^{0})$ with the
 true value of $\alpha=0$, where    $H$ is defined in Theorem 1. Then  $\tilde{\bx}_i^{T}\beta = \tilde{\bx}_i^{T}\beta^{0}+U_i^{T}\alpha$, where $U_{i} = H^{-1}\tilde{\bx}_{i}$.  The local likelihood (\ref{local log-like}) is rewritten as
\begin{eqnarray}
\ell_{x}  =  \frac{1}{n} \sum^{n}_{i=1}\{\Delta_{i}(\tilde{\bx}^{T}_{i}\beta^{0}+U^{T}_{i}\alpha) + \Delta_{i} \log f(Y_{i}; \gamma_0)
   - \exp(\tilde{\bx}^{T}_{i}\beta^{0}+U^{T}_{i}\alpha)F(Y_{i}; \gamma_0)\}K_{h}(X_{i}-x). 
\label{lx2}
 \end{eqnarray}
Taking the derivative of (\ref{lx2}) with respect to $\alpha$ yields
\begin{equation}
  \frac{1}{n} \sum^{n}_{i=1}\{\Delta_{i} - \exp({\bf \tilde{x}}^{T}_{i}\beta^{0}+U^{T}_{i}\alpha)F(Y_{i};\gamma_{0})\}U_{i}K_{h}(X_{i}-x).
\label{lx3}
\end{equation}
Then it is equivalent to show that there exists a maximizer $\hat{\alpha}$ to the likelihood equation
(\ref{lx3}) such that  $\hat{\alpha}  \stackrel{p}{\rightarrow} 0$.

Let $B(0,\epsilon)$ be an open ball which centered at $0$ with radius $\epsilon > 0$. Denote
by $\alpha_j$ the $j$-th element of $\alpha$.
By a Taylor expansion around the true $\alpha=0$,
$\ell_{x}(\alpha) = \ell_{x}(0) + \ell^{'}_{x}(0)^{T}\alpha + \frac{1}{2}\alpha^{T}\ell^{''}_{x}(0)\alpha + R_{n}(\alpha^{\star}),$
where $R_{n}(\alpha) = \frac{1}{3!}\sum_{j,k,l} \alpha_{j} \alpha_{k} \alpha_{l} 
\frac{\partial^{3}}{\partial\alpha_{j}\partial\alpha_{k}\partial\alpha_{l}}\ell_{x}(\alpha)$ and $\alpha^{*}$ is between $\alpha$ and 0.  For the term $ \ell^{'}_{x}(0)^{T}\alpha$,  
\[
 \ell^{'}_{x}(0)  =   \frac{1}{n} \sum^{n}_{i=1}\{\Delta_{i} - \exp(X^{T}_{i}\beta^{0})F(Y_{i};\gamma_{0})\}U_{i}K_{h}(X_{i}-x) 
     \stackrel{p}{\rightarrow}  E\{\Delta - \theta(x)F(Y;\gamma_{0})|X=x\}\int \begin{bf}u\end{bf} K(u)du, 
\]
where  $\bu$ is  defined in Theorem 2. Based on Lemma 1, $ \ell^{'}_{x}(0)=0$.
Thus  for any $\epsilon > 0$, with probability tending to 1, $$| \ell^{'}_{x}(0)^{T} \alpha | \leq \epsilon\ \ \mbox{for } \alpha \in B(0,\epsilon).$$

For $ \alpha^{T}\ell^{''}_{x}(0) \alpha$,
taking the second derivative of (\ref{lx2}) with respect to $\alpha$ yields  $\ell_x^{''}$,
\begin{eqnarray}
  \ell^{''}_{x} & = &  \frac{1}{n} \sum^{n}_{i=1}\{-\exp({\bf \tilde{x}}^{T}_{i}\beta^{0}+U^{T}_{i}\alpha)F(Y_{i};\gamma_{0})\}U_{i}U^{T}_{i}K_{h}(X_{i}-x). \nonumber
\end{eqnarray}
Plugging in  $\alpha=0$, the expectation of  $\ell_x^{''}(0)$ is 
\begin{eqnarray}
   & &   -E\{\exp({\bf \tilde{x}}^{T}\beta^{0})F(Y;\gamma_{0})UU^{T}K_{h}(X-x)\} \nonumber \\ 
    & = & -\int \int \exp(m(X))F(Y;\gamma_{0})UU^{T}K_{h}(X-x)f_{Y|X}(Y|X) f_X(X)dXdY  (1+ o_P(1)) \nonumber\\ 
  & = & -\int \int \exp(m(x+hu))F(Y;\gamma_{0})\begin{bf}uu^{T}\end{bf}K(u)f_{Y|X}(Y|X = x)f_X(x+hu)dudY  (1+ o_P(1))  \nonumber\\ 
  & =& -f_X(x)\int E(\theta(x)F(Y;\gamma_{0})|X=x) \begin{bf}uu^{T}\end{bf}K(u)du  (1+ o_P(1))  \nonumber\\ 
 & = &  -f_X(x) E(\Delta |X=x) \int  \begin{bf}uu^{T}\end{bf}K(u)du  (1+ o_P(1))= -f_X(x)S_{1}(x)  (1+ o_P(1)),
\label{lx6}
\end{eqnarray}
where $ S_{1}(x)$ is defined in Theorem 2. 
Thus  for any  $\epsilon > 0 $, $$\alpha^{T}\ell^{''}_{x}(0) \alpha < -f_X(x) \eta_{1} \epsilon^{2} \mbox{ with probability tending to 1},$$
where $\eta_{1}$ is the minimum eigenvalue of $S_{1}(x)$ (Horn {\it et al.} (1998)).

Under condition (A9), 
 $|R_{n}(\alpha)| \leq C\epsilon^{3}\frac{1}{n}\sum^{n}_{i=1}M(Y_{i})=C\epsilon^{3}\{E(M(Y))+o_{p}(1)\}$
for some constant $C > 0$. As a result, when $\epsilon$ is small enough, for any $\alpha \in B(0,\epsilon)$,
$$\ell_{x}(\alpha) - \ell_{x}(0) \leq 0 \mbox{\ with probability tending to 1},$$ which implies
$$\sup_{\alpha\in B(0, \epsilon)} \ell_{x}(\alpha) - \ell_{x}(0) \leq 0 \mbox{\ with probability tending to 1}.$$
Thus  $\ell_x(\alpha)$ has a local maximum in $B(0, \epsilon)$ so that   the likelihood equation  has a maximizer
$\hat{\alpha}(\epsilon)$ and 
$||\hat{\alpha}|| \leq \epsilon$ with probability tending to 1.
This completes the proof of Theorem 1.

\begin{bf}
\noindent Proof of Theorem 2\\
\end{bf}
 We prove part (a) first. Following the proof of Theorem 1, 
from (\ref{lx2}) with $\gamma=\gamma_0$,  since
$0 = \ell_x^{'}(\hat{\alpha})  \approx  \ell_x^{'}(0) +  \ell_x^{''}(0)\hat{\alpha},$
\begin{equation}
\hat{\alpha} \approx - \{\ell_x^{''}(0) (1+ o_P(1))\}^{-1} \ell_x^{'}(0).
\label{lx1}
\end{equation}
We derive the asymptotic expressions of $\ell_x^{'}(0)$  and $Var\{\ell_x^{'}(0)\}$.

Taking the expectation of (\ref{lx3}) and using Lemma 1,
\begin{eqnarray}
 & &   E\{ (\Delta - \exp({\bf \tilde{x}}^{T}\beta^{0})F(Y;\gamma_{0})) UK_{h}(X-x)\} \nonumber\\
    & &   = E_{X}\{E_{Y|X}\{\Delta - \exp({\bf \tilde{x}}^{T}\beta^{0} )F(Y;\gamma_{0}) \} U K_{h}(X-x)\} \nonumber\\
     & &   = E_{X}\{E_{Y|X}\{F(Y;\gamma_{0}) \exp(m(X))  - \exp({\bf \tilde{x}}^{T}\beta^{0} ) F(Y;\gamma_{0}) \} U K_{h}(X-x)\} \nonumber\\
 & &=  E_{X}\{E_{Y|X}\{ F(Y;\gamma_{0})[ \exp(m(X)) - \exp({\bf \tilde{\bx}}^{T}\beta^{0})] \} U K_{h}(X-x) \}.
\label{lx4}
  \end{eqnarray}  
By a Taylor expansion, 
\[ \exp(m(X)) - \exp({\bf \tilde{\bx}}^{T}\beta^{0})= \theta(x)\frac{m^{(p+1)}(x)}{(p+1)!}(X-x)^{p+1} (1 + o_{p}(1)),
\]
and by a change of variables $X-x = hu$, 
 (\ref{lx4})  is
 \begin{eqnarray} 
    & &   \int\int  \theta(x)\frac{m^{(p+1)}(x)}{(p+1)!}(hu)^{p+1}F(Y;\gamma_{0})\begin{bf}u\end{bf} K(u)f_{Y|X}(Y|X = x)f_X(x+hu)du dY    (1 + o_{p}(1)) \nonumber \\   
   & & =   f_X(x)\theta(x)\frac{m^{(p+1)}(x)}{(p+1)!}h^{p+1}\int \int F(Y;\gamma_{0})f_{Y|X}(Y|X = x)dY  u^{p+1}\begin{bf}u\end{bf} K(u)du  (1 + o_{p}(1))   \nonumber \\  
    & &  = f_X(x)\theta(x)\frac{m^{(p+1)}(x)}{(p+1)!}h^{p+1} E\{F(Y;\gamma_{0})|X = x\} \int  \begin{bf}u\end{bf}u^{p+1}K(u)du   (1 + o_{p}(1)).\nonumber 
          \end{eqnarray} 
Applying Lemma 1 again,  the last expression is 
 \begin{equation}
 f_X(x)\frac{m^{(p+1)}(x)}{(p+1)!}h^{p+1} E\{\Delta |X = x\} \int  \begin{bf}u\end{bf}u^{p+1}K(u)du (1+ o_P(1)). 
\label{lx5}
\end{equation}
By (\ref{lx6}), (\ref{lx1}), and (\ref{lx5}),    $b_n(x)$ is obtained.

For 
$Var\{\ell_x^{'}(0)\}$, it can be decomposed into two parts,  the   quadratic term, and  the cross-product terms minus the squared of $E\{\ell_x^{'}(0)\}$. For the quadratic term, 
\begin{eqnarray}
& & \frac{1}{n^{2}} E \left\{ \sum^{n}_{i=1}K^{2}_{h}(X_{i}-x)
 ([\Delta_{i} - \exp({\bf \tilde{x}}^{T}_{i}\beta^{0})F(Y_{i};\gamma_{0})]U_{i} )([\Delta_{i} - \exp({\bf \tilde{x}}^{T}_{i}\beta^{0})F(Y_{i};\gamma_{0})]U_{i})^{T} \right\} \nonumber  \\
& = & \frac{1}{n}E\left\{ K^{2}_{h}(X-x) ([\Delta - \exp({\bf \tilde{x}}^{T} \beta^{0})F(Y; \gamma_{0})]U)([\Delta -  \exp({\bf \tilde{x}}^{T} \beta^{0})F(Y; \gamma_{0})]U )^{T} \right\} \nonumber  \\
& = & \frac{1}{nh}f_X(x)\int K^{2}(u) ([\Delta - \theta(x)F(Y_{i};\gamma_{0})])^{2}\begin{bf}u\end{bf}\begin{bf}u\end{bf} )^{T}du (1+o_p(1))
 = \frac{1}{nh}f_X(x)S_{2}(x; \gamma_{0}) (1+o_p(1)),\nonumber  
\end{eqnarray}  
where $S_{2}(x; \gamma_{0})$ is defined in Theorem 2.
Based on (\ref{lx5}),  the cross-product terms minus the squared of $E\{\ell_x^{'}(\beta^0)\}$ is of order $n^{-1}h^{2(p+1)} (1+o_p(1))$.
Hence  
\begin{equation}
Var\{\ell_x^{'}(0)\} = 
(nh)^{-1} f_X(x)S_{2}(x; \gamma_{0})  (1+o_p(1)).
\label{lx7}
\end{equation}
Based on  (\ref{lx6}), (\ref{lx1}), and (\ref{lx7}),  $Var(H \hat{\beta})$ in Theorem 2 is obtained.

To prove asymptotic normality, by the Cramer-Wold device,  for any non-zero constant vector 
$b \in R^{p+1} $,  we will show that 
$$\sqrt{nh}\{b^{T} H (\hat{\beta}-\beta) - b^{T} b_n(x)\} \rightarrow N (0, f_X(x)^{-1}b^T S^{-1}_{1}(x) S_{2}(x;\gamma_{0}) S^{-1}_{1}(x)b).$$
Consider $\sqrt{nh}\{b^{T} \ell^{'}_x(0)-  b^{T}E(\ell^{'}_x(0))\}$ at first. We verify the Lyapounov condition as follows. For $0< \eta <1$, 
\begin{eqnarray*}
 & &  \sum^{n}_{i=1} E\{ |\sqrt{nh}n^{-1}b^{T} ([\Delta_{i} - \exp(\tilde{\bx}^{T}_{i}\beta^{0})F(Y_{i};\gamma_{0})]U_{i}K_{h}(X_{i}-x)  \\  & & \mbox{ } -E ([\Delta_{i} -
\exp(\tilde{\bx}^{T}_{i}\beta^{0})F(Y_{i};\gamma_{0})]U_{i}K_{h}(X_{i}-x)) ) |^{2+\eta}\}\\
  &  \leq &  2(n^{-1}h)^{1+\frac{\eta}{2}}n b^{T}E\{| [\Delta - \exp(\tilde{\bx}^{T} \beta^{0})F(Y; \gamma_{0})]U K_{h}(X -x)|^{2+\eta}\}\\
 & \leq &  2( (nh)^{\frac{-\eta}{2}}h^{1+\eta} b^{T}E\{| [\Delta - \exp(\tilde{\bx}^{T} \beta^{0})F(Y;\gamma_{0})]U K_{h}(X-x)|^{2+\eta}\}\\
 &  = &  O((nh)^{\frac{-\eta}{2}}h^{1+\eta} ) \rightarrow 0.
\end{eqnarray*}
Then  the asymptotic normality of   $\sqrt{nh}\{b^{T} \ell^{'}_{x}(0)-  b^{T} E( \ell^{'}_{x}(0))\}$ is shown;
that is,
\begin{equation}
\sqrt{nh}\{ \ell^{'}_{x}(0)-  E( \ell^{'}_{x}(0))\} \rightarrow N(0, f_X(x) S_{2}(x;\gamma_{0})).
\label{lx8}
\end{equation}
Finally,  the dominant term of  $\sqrt{nh}\{H(\hat{\beta}-\beta) - b_n(x)\}$ is $\{f_X(x)S_{1}(x)\}^{-1}\{ \ell^{'}_{x}(\beta^{0})- E( \ell^{'}_{x}(\beta^{0})) \}$, and with  (\ref{lx8}), Theorem 2(a) is proved.
 
 To show Theorem 2(b), note that the results in (a) depend on $\gamma_0$ only through $S_2(x; \gamma_0)$.
From the definition of $S_2(x; \gamma_0)$,  it is clear that  if $\hat{\gamma}$ is $\sqrt{n}$-consistent, then  $F(Y; \hat{\gamma})$ is $\sqrt{n}$-consistent and the results in (\ref{lx7}) and (\ref{lx8}) continue to hold.

\begin{bf}
\noindent Proof of Theorem 3\\
\end{bf}
%
%
%
\noindent (a) Given that $\theta(\cdot)$ is known,  the poof is similar to proofs for standard maximum likelihood estimators. 
First,
\begin{equation}
0 = \ell^{*'}(\hat{\gamma}) \approx  \ell^{*'}(\gamma_{0}) +  \ell^{*''}(\gamma_{0})(\hat{\gamma}-\gamma_{0}).
\label{lx9}
\end{equation}
Then $\sqrt{n}  (\hat{\gamma}-\gamma_{0}) \approx -\sqrt{n}(\ell^{*''}(\gamma_{0}))^{-1} \ell^{*'}(\gamma_{0}).$ 
From (\ref{noncured}),  the first derivative of $\ell^{*}$ with respect to $\gamma$ is
\begin{eqnarray}
 n^{-1}\sum^{n}_{i=1}I(Y_{i} < \infty)\left\{\Delta_{i}[\xi^{'}(Y_{i};\gamma)-\theta(X_{i})F^{'}(Y_{i};\gamma)] - (1-\Delta_{i})\left[\frac{\theta(X_{i})F^{'}(Y_{i};\gamma)}{1 - \exp(-\theta(X_i)\bar{F}(Y_{i};\gamma))}\right]\right\}, \nonumber \\
\label{lstar1}
\end{eqnarray}
where   $\xi(Y; \gamma)$ is defined in condition (A8), $\bar{F}(\cdot)= 1-F(\cdot)$,  and $F^{'}(\cdot)$ and $\xi^{'}(\cdot)$ are the first derivatives  of $F$ and $\xi$ respectively  with respect to $\gamma$. 
Since $E\{  \ell^{*'}(\gamma_{0}) \}=0$, the bias of $\hat{\gamma}$ is 0. 
We know that $\ell^{*''}(\gamma_{0}) = {n}^{-1}\sum^{n}_{i=1}\ell^{*''}_i(\gamma_{0}) \rightarrow_{p}  E\{ \ell^{*''}_i(\gamma_{0})\}\equiv T_1(\gamma_0)$. The expression of $T_1(\gamma)$ is obtained by deriving  $\frac{\partial^{2}\ell^{*}}{\partial\gamma^{2} }$ 
from (\ref{lstar1}) and then taking expectation, 
\begin{eqnarray}
 T_1(\gamma)  & = & E \left\{
I(Y < \infty)\left\{\Delta [ \xi^{''}(Y; \gamma) - \theta(X)F^{''}(Y;\gamma)]
+(1- \Delta) \frac{\theta(X)} {1 - \exp(-\theta(X)\bar{F}(Y; \gamma))}  \right.  \right. \nonumber \\
  & & \hspace{0.5cm}  \times \left. \left. \left[ F^{''}(Y; \gamma)+ \frac{F^{'^2}(Y; \gamma) \exp(-\theta(X) \bar{F}(Y;\gamma))} {1 - \exp(-\theta(X)\bar{F}(Y; \gamma))} \right] \right\}
 \right\}.
\label{t1}
\end{eqnarray}
For
$Var(\hat{\gamma}-\gamma_{0})$, based on (\ref{lx9}), it is approximately $T_{1}(\gamma_{0})^{-1} E\{(\ell^{*'}(\gamma_0))^2 \} T_{1}(\gamma_{0})^{-1}. $
For $E\{(\ell^{*'}(\gamma_{0}))^2\}$, it is dominated by  $n^{-2}\sum^{n}_{i=1}E\{  (\ell_i^{*'}(\gamma_{0}))^2\} \equiv n^{-1}  T_{2}(\gamma_0),$
where 
\begin{equation} T_{2}(\gamma)= E \left\{
I(Y < \infty)\left\{\Delta [\xi^{'}(Y ; \gamma)-\theta(X)F^{'}(Y; \gamma)] - (1-\Delta )\left[\frac{\theta(X)F^{'}(Y ; \gamma)}{1 - \exp(-\theta(X)\bar{F}(Y;\gamma))}\right]\right\}^2
\right\}.
\label{t2}
\end{equation} 
Moreover $  \sqrt{n}\ell^{*'}(\gamma_{0})$ converges in distribution to $N(0,  T_2(\gamma_0))$. By Slutsky's theorem,  Theorem 3(a) is proved.\\
(b) 
When $\theta(\cdot)$ is estimated at the rate in Theorem 2,  the difference between the  expected values of (\ref{lstar1}) with
true $\theta(x)$ and 
$\hat{\theta}(x)$ for $x$ evaluated at a data point $X_{i}$ is of rate $h^{(p+1)}$. The conditions $nh^{2p+2} \rightarrow 0$ and  $nh \rightarrow \infty$  ensure that  the asymptotic normality in (a) continues to hold.

\renewcommand\bibname{Reference}

 \vspace{1cm} 

\noindent {\bf Acknowledgment:} The authors were partially supported by the  Ministry of Science and Technology in Taiwan under Grant MOST 103-2118-M-007-001-MY2.  The second author wishes to dedicate this paper to the memory
of  Professor Andrei Yakovlev.
\vspace{0.5cm}


\begin{table}[h]
\captionof{table}{Summary of simulation results in Example 1
}
\[\begin{tabular}{c|cccc|cc}
			\toprule[2pt]  
$h$ &  \multicolumn{4}{c|}{performance of $\hat{\gamma}$}  &   \multicolumn{2}{c}{MSE ($\hat{m}$) (sd)} \\
 &   mean($\hat{\gamma}$) &  sd of $\hat{\gamma}$ & $\widehat{se}(\hat{\gamma})$  & 95\% coverage rate &  $\gamma$ known & with $\hat{\gamma}$\\
			 \midrule[1pt] 
0.2  & 6.879 & 0.924 & 0.867 & 91.2$\%$ & 0.078 (0.041)  & 0.084 (0.043)\\
0.4  & 7.127 & 0.940 & 0.900 & 93.1$\%$ & 0.035 (0.023)& 0.039 (0.025) \\
0.6  & 7.142 & 0.957 & 0.903 & 92.8$\%$ & 0.025 (0.018)  & 0.029 (0.022)\\
  \bottomrule[2pt]
\end{tabular}\]
\end{table}

\begin{table}[h]
\captionof{table}{Summary of simulation results in Example 2
}
\[\begin{tabular}{c|cccc|cc}
			\toprule[2pt]  
$h$ &  \multicolumn{4}{c|}{performance of $\hat{\gamma}$}  &   \multicolumn{2}{c}{MSE ($\hat{m}$) (sd)} \\
 &   mean($\hat{\gamma}$) &  sd of $\hat{\gamma}$ & $\widehat{se}(\hat{\gamma})$  & 95\% coverage rate &  $\gamma$ known & with $\hat{\gamma}$\\
			 \midrule[1pt]
0.2  & 6.974 & 0.840 & 1.165 & 96.9 $\%$ &  0.204 (0.298) & 0.205 (0.299)\\
0.4 & 7.116 & 0.849 & 1.194  & 97.0 $\%$ &  0.075 (0.069) & 0.075 (0.069) \\
0.6  & 7.152 & 0.853 & 1.192  & 97.0 $\%$ &  0.047 (0.042)  & 0.048 (0.042)\\
  \bottomrule[2pt]
\end{tabular}\]
\end{table}

\newpage

\begin{figure}[h]  
 \centering
 \caption{Example 1 with $h=0.6$:  (a) the true (solid line) and estimated  curves of $m(\cdot)$ with 10th (long dashed line),  50th (dashed line), and  90th percentiles (short dashed line) of MSE; (b) the corresponding cure rate curves; (c)
  estimated $m(\cdot)$  (dashed line) with 50-th percentile of  MSE and its 95\% pointwise confidence intervals (dotted line) when $\gamma$ known;  and (d) $\gamma$ estimated.
Points show the observed response, cross (failure), circle (censored), and triangle (cured) for the data set with 50-th percentile of average MSE.} \label{E1_200_06}
\includegraphics[width=16 cm,height=16 cm]{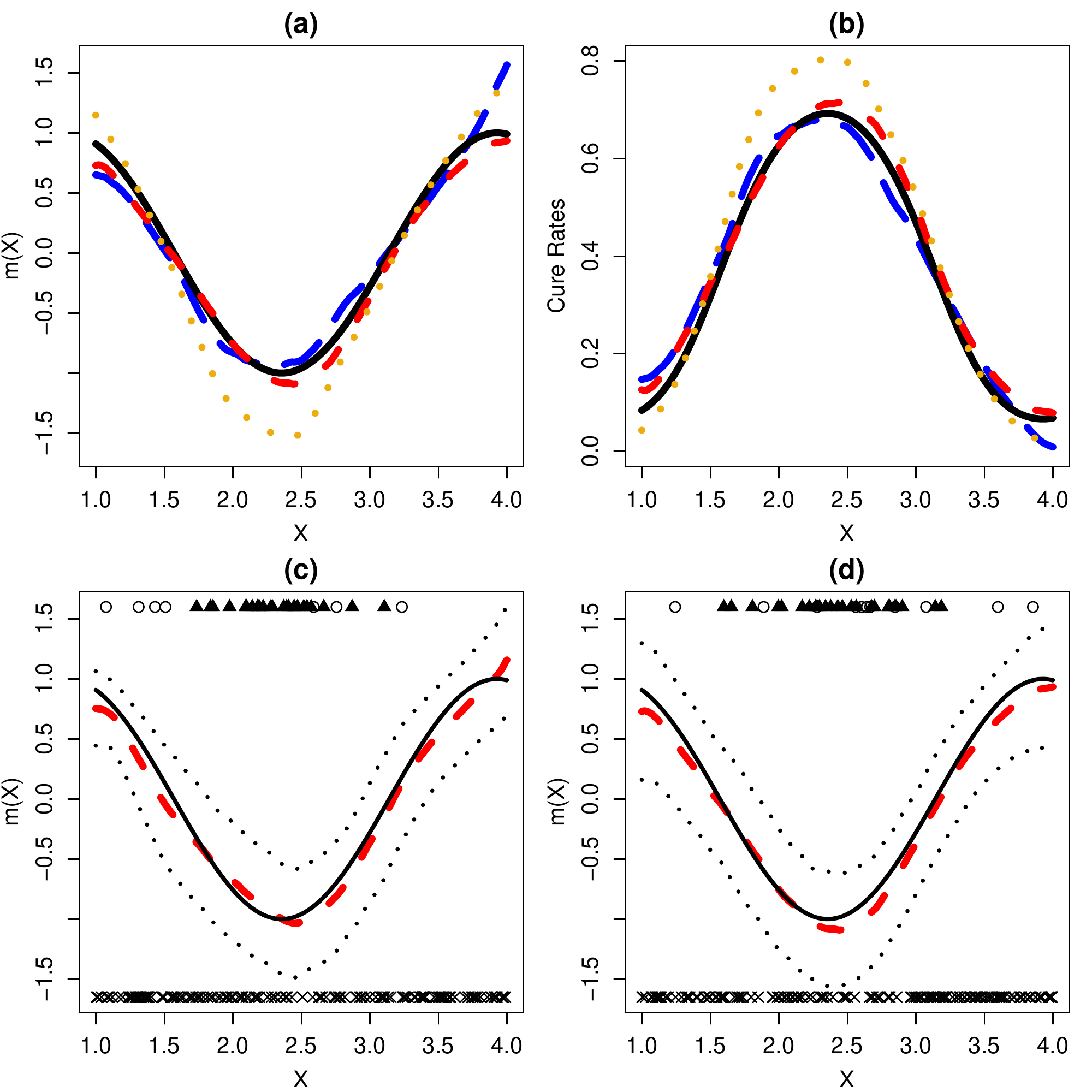}
\end{figure} 

\newpage

\begin{figure}[h]  
 \centering
 \caption{Example 2 with $h=0.6$:  (a) the true (solid line) and estimated  curves of $m(\cdot)$ with 10th (long dashed line),  50th (dashed line), and  90th percentiles (short dashed line) of MSE; (b) the corresponding cure rate curves; (c)
  estimated $m(\cdot)$  (dashed line) with 50-th percentile of  MSE and its 95\% pointwise confidence intervals (dotted line) when $\gamma$ known;  and (d) $\gamma$ estimated.
Points show the observed response, cross (failure), circle (censored), and triangle (cured) for the data set with 50-th percentile of average MSE.} \label{E2_200_06}
\includegraphics[width=16 cm,height=16 cm]{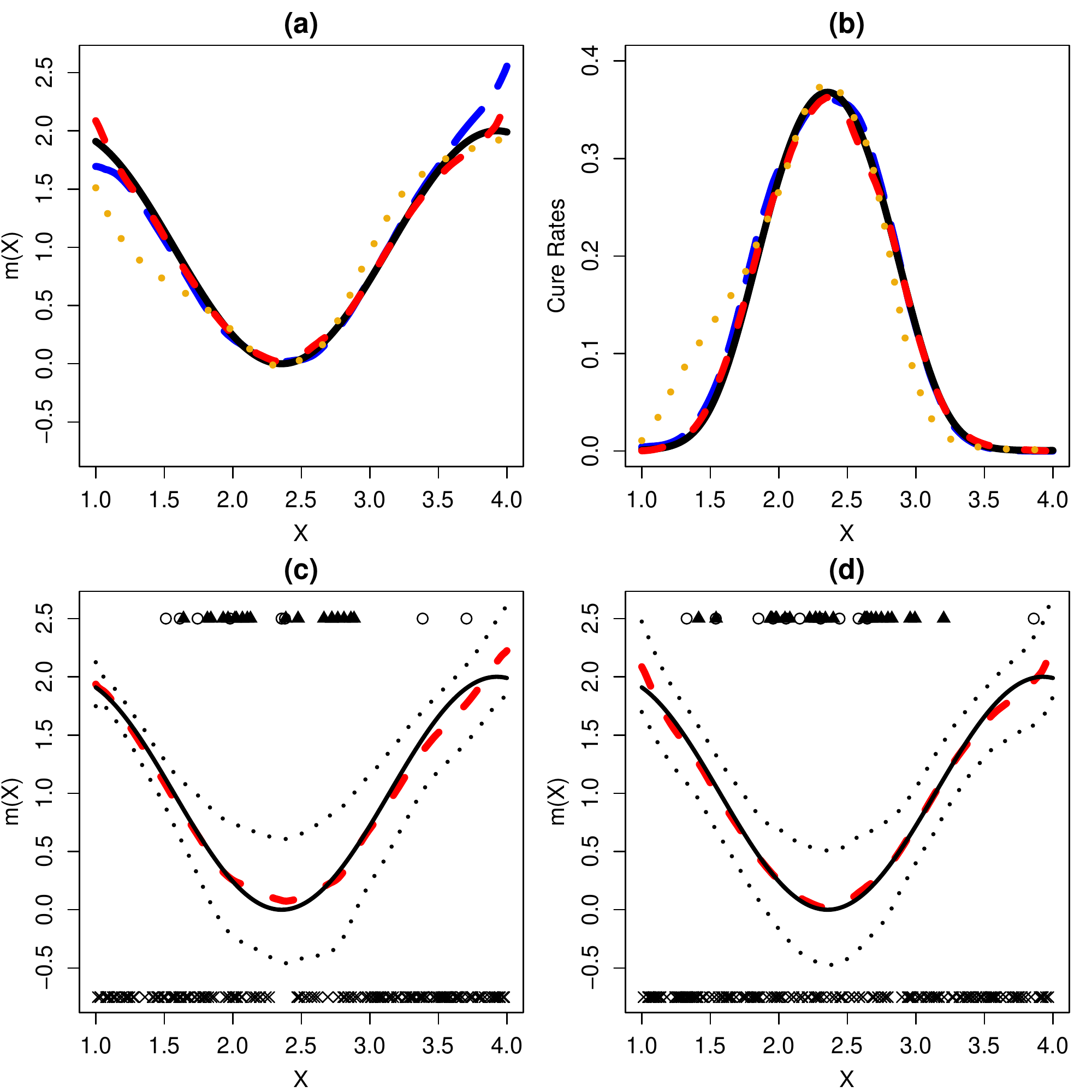}
\end{figure} 

\newpage
 \begin{figure}[h]  
 \centering
 \caption{Kidney Transplant  Data: (a) the estimated cure rates with $h=10$ in step 2 and $h=$18 (red line) and 22  (blue line)  in step 5, and their  95$\%$ pointwise confidence intervals (dotted lines); (b) the estimated cure rates with $h=12$ in step 2 and $h=$18 (blue line) and 22 (red line)  in step 5, and their  95$\%$ pointwise confidence intervals (dotted lines). Panel descriptions for (c) and (d) are as for (a) and (b) except that the curves plotted are $\hat{m}(\cdot)$. 
Points show the observed response, cross (failure), circle (censored), and triangle (cured).
} \label{R2}
\includegraphics[width=16 cm,height=16 cm]{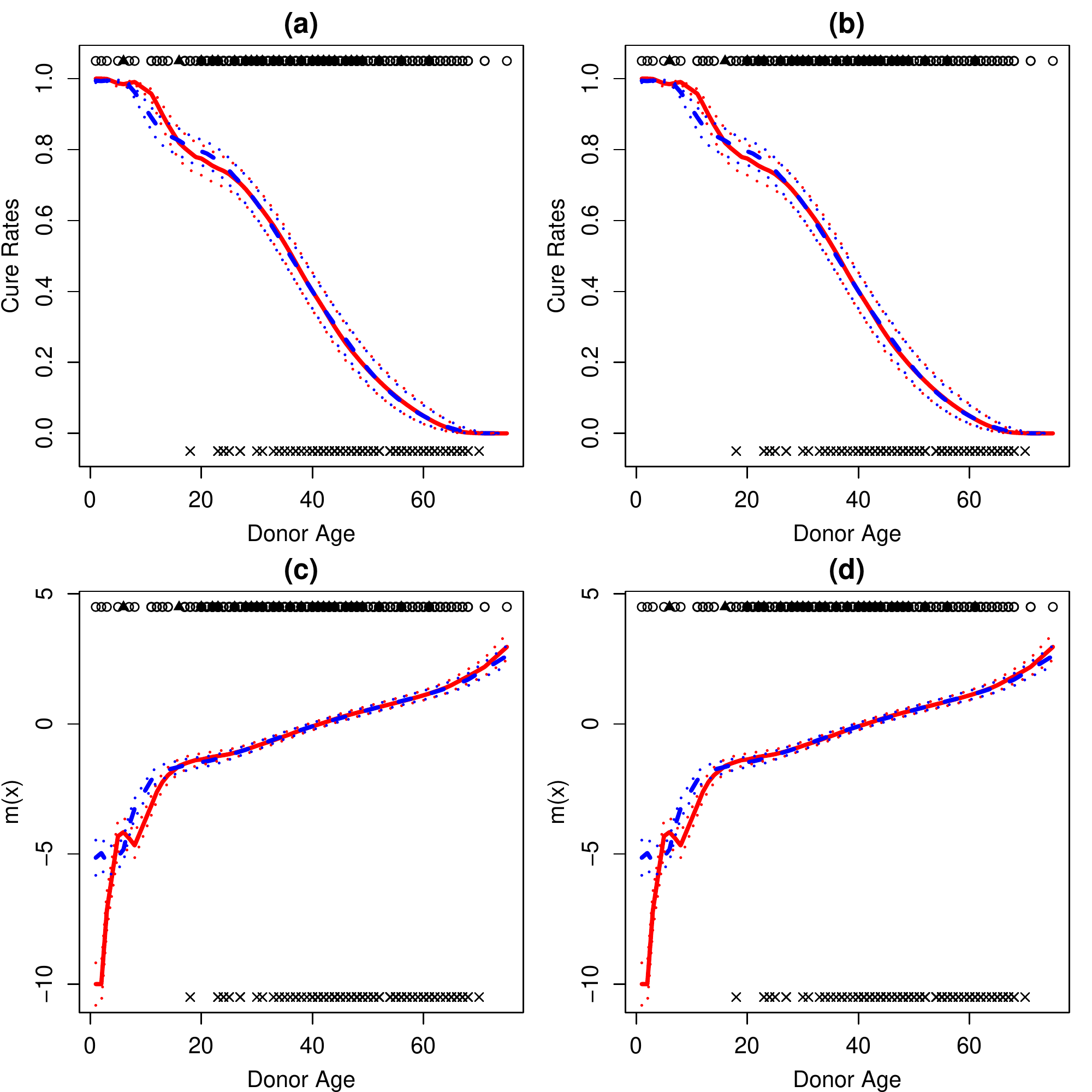}
\end{figure}

\begin{figure}[h]  
 \centering
 \caption{Kidney Transplant  Data: (a) the estimated cure rates and their 95\% pointwise confidence intervals (dotted lines)  with $h=10$ in step 2 and $h=$22 in step 5 using  cure thresholds  3100 (black line) and 3147 (blue line); (b) using  cure thresholds 3200 (black line) and 3300 (red line); (c) the estimated survival curve (solid line) under model (1) with the empirical Kaplan-Meier curve (dashed line); (d) estimated hazard functions  under model (1) conditioned on  patient age, first quartile 33 (solid line), mean 42.84 (long dashed line), and 3rd quartile 54 (dashed line) years. Points in (a)(b) show the observed response, cross (failure), circle (censored), and triangle (cured).
} \label{R2_diffc}
\includegraphics[width=16 cm,height=16 cm]{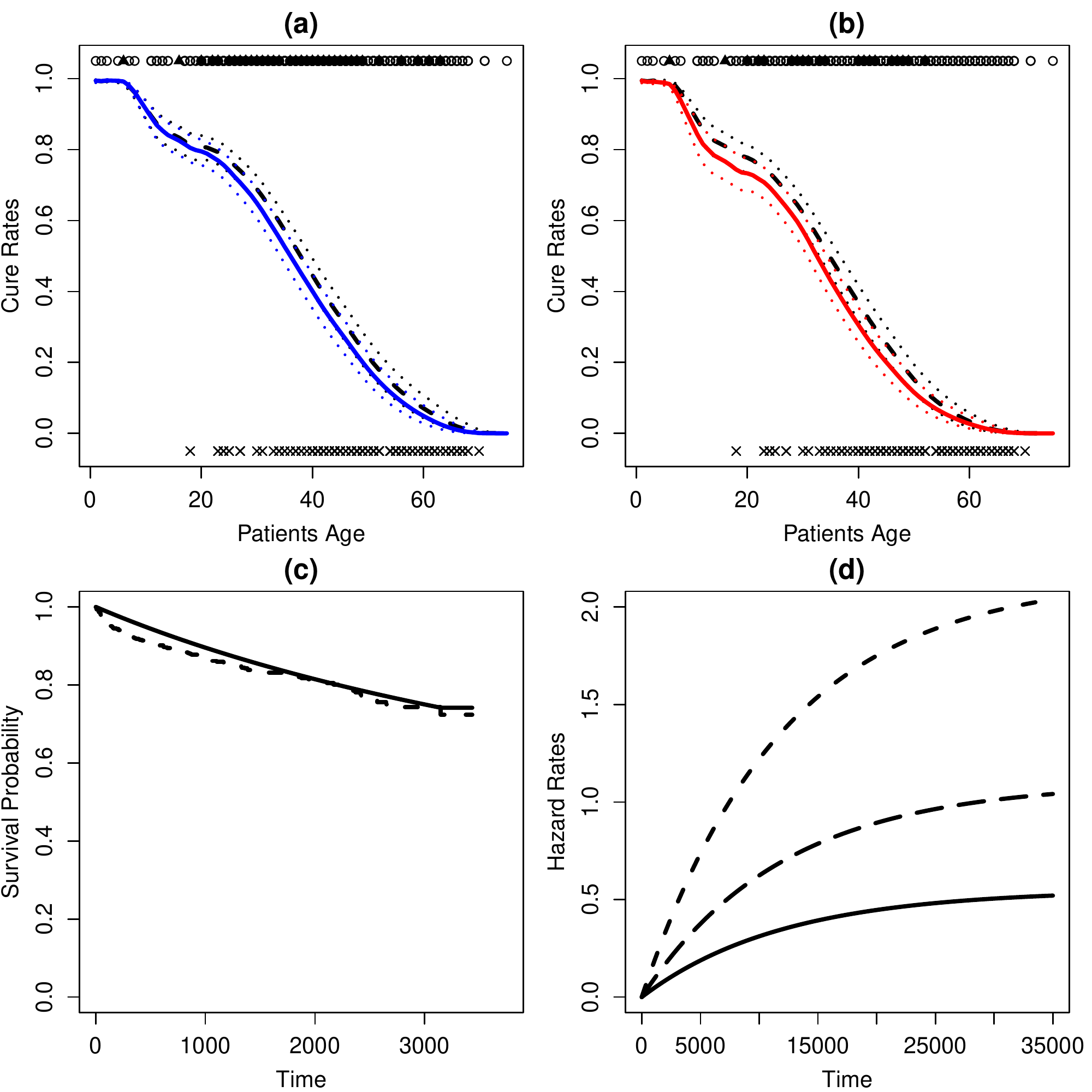}
\end{figure}


\begin{thebibliography}{9}  
\bibitem{ingrid}
Bertrand, A., Legrand, C.,  Carroll, R. J., de Meester,  C., and Van Keilegom, I. (2017), ``Inference in a Survival Cure Model with Mismeasured Covariates using a Simulation-Extrapolation Approach," \emph{Biometrika}, 104, 31-50.

\bibitem{TwoCCureM}
Berkson, J. and Gage, R. P. (1952), ``Survival Curve for Cancer Patients Following Treatment," \emph{Journal of the American Statistical Association}, 47, 501-515.

\bibitem{cai}
Cai, J., Fan, J., Jiang, J., and Zhou, H. (2007),  ``Partially Linear Hazard Regression for Multivariate Survival Data," \emph{Journal of the American Statistical Association}, 102, 538-551.

\bibitem{chen}
Chen, M. H., Ibrahim, J. G., and Sinha, D. (1999),  ``A New Bayesian Model for Survival Data with a Surviving Fraction," \emph{ Journal of the American Statistical Association}, 94, 909-919.

\bibitem{chendu}
Chen T., and  Du P. (2018), ``Promotion Time Cure Rate Model with Nonparametric Form of Covariate
Effects," \emph{Statistics in Medicine}, 37, 1625–1635.



\bibitem{Cox}
Cox, D. R. (1975), ``Partial likelihood," \emph{ Biometrika}, 62, 269-276.

\bibitem{Fan1996}
Fan, J. and Gijbels, I. (1996), {\it Local Polynomial Modelling and Its Applications}, London: Chapman and Hall.

\bibitem{Fan1997}
Fan,  J., Gijbels, I., and King, M. (1997), ``Local Likelihood and Local Partial Likelihood in Hazard Regression," \emph{The Annals of Statistics}, 25,  1661-1690.


\bibitem{Fare1982}
Farewell, V. (1982), ``The Use of Mixture Models for the Analysis of
Survival Data with Long-term Survivors,"  \emph{Biometrics,} 38, 1041-1046.

\bibitem{hanin}
Hanin, L. and Huang, L.-S. (2014),  ``Identifiability of Cure Models Revisited," {\it Journal of Multivariate Analysis}, 130, 261-274.

\bibitem{hosmer2008}
Hosmer, D. W., Lemeshow, S. and May, S. (2008), {\it Applied Survival Analysis: Regression Modeling of Time to Event Data}, New York: John Wiley and Sons Inc.

\bibitem{horn1998}
Horn, R. A., Rhee, N. H., and So, W. (1998), ``Eigenvalue inequalities and equalities," \emph{Linear Algebra and its Applications}, 279, 29-44.

\bibitem{huang1999}
Huang, J. (1999), ``Efficient Estimation of the Partly Linear Additive Cox Model," \emph{ The Annals of Statistics}, 27, 1536-1563.


\bibitem{Book.Sur}
Klein, J. P. and Moeschberger, M. L.  (1997), \emph{Survival Analysis Techniques for Censored and Truncated Data}, New York: Springer.

\bibitem{Kuk1992}
Kuk, A. Y. C. and Chen, C. H. (1992), ``A Mixture Model Combining Logistic Regression with Proportional Hazards Regression," \emph{Biometrika}, 79, 531-541.

\bibitem{li2001}
Li, C.-S., Taylor, J. M. G., and Sy, J. P. (2001), ``Identifiability of Cure Models," \emph{Statistics and Probability Letters,} 54,  389-395.

\bibitem{La1992}
Laska, E. M. and Meisner, M. J. (1992), ``Nonparametric estimation and testing in a cure model." \emph{Biometrics}, 48, 1223-1234.

\bibitem{Lin2015}
Lin, L.-H. (2015), Cure Rate Models with Local Polynomial Estimation (Master Thesis),  National Tsing-Hua University, Hsinchu, Taiwan.


\bibitem{Lu2004}
Lu, W. and Ying, Z. (2004), ``On Semiparametric Transformation Cure
Models,"  \emph{Biometrika}, 91, 331-343.

\bibitem{mayin}
Ma, Y. and Yin, G.  (2008), ``Cure Rate Model with Mismeasured Covariates under Transformation," \emph{Journal of the American Statistical Association}, 103,  743-756.

\bibitem{maowang}
Mao, M. and Wang, J.  L. (2010), ``Semiparametric Efficient Estimation for a Class of Generalized Proportional Odds Cure Models," \emph{Journal of the American Statistical Association}, 105, 302-311.

\bibitem{Tish1987} 
Tibshirani,  R.  and  Hastie, T. (1987), ``Local Likelihood Estimation," \emph{Journal of the American Statistical Association}, 82, 559-567.

\bibitem{T1998}
Tsodikov, A. (1998), ``A Proportional Hazards Model Taking Account of Long-Term  Survivors". \emph{Biometrics}, 54, 1508-1516.

\bibitem{T2003}
Tsodikov, A. D., Ibrahim, J. G., and Yakovlev, A. Y. (2003), ``Estimating Cure Rates from Survival Data: an Alternative to Two-Component Mixture Models, "\emph{Journal of the American Statistical Association}, 98, 1063-1078.

\bibitem{wang2012}
Wang, L., Du, P., and Liang, H. (2012), ``Two-Component Mixture Cure Rate Model with Spline Estimated Nonparametric Components," \emph{Biometrics}, 68, 726-735.

\bibitem{YT1996}
Yakovlev, A. Y. and Tsodikov, A.D. (1996),  \emph{Stochastic Models of Tumor Latency and Their Biostatistical Applications}, Singapore: World Scientific.

\bibitem{Zeng2006}
Zeng,  D.,  Yin,  G., and Ibrahim,  J. G. (2006), ``Semiparametric Transformation Models for Survival Data with a Cure Fraction," \emph{Journal of the American Statistical Association}, 101, 670-684.
\end{thebibliography}
\end{document}